%% file: article.tex
\documentclass[12pt]{article}
\usepackage{epsfig}
\usepackage{graphicx}
\usepackage{subfigure}

\input econfmacros.tex
\def\Title#1{\begin{center} {\Large {\bf #1} } \end{center}}

\begin{document}

\Title{Interior matter estimates of rapidly rotating compact
stars}

\bigskip\bigskip


\begin{raggedright}

{\it Nana Pan\index{Pan, NN.}}\\
College of Mathematics and Physics \\
Chongqing University of Posts and Telecommunications \\
Chongqing 400065\\
P. R. China\\
{\tt Email: cqpannn@gmail.com}\\
{\it Xiaoping Zheng \index{Zheng, XP.}}\\
Institute of Astrophysics \\
Huazhong Normal University \\
Wuhan 430079\\
P. R. China\\

\bigskip\bigskip
\end{raggedright}

\section{Introduction}

The components and properties of interiors in neutron stars have
attracted much attention since the discovery of the first pulsar
and then the confirmation of it to be a fast rotational neutron
star. However, the composition of matter for neutron star at
super-nuclear density is still an indeterminacy due to the
uncertain nuclear physics. This is also the main central question
for the astronomy and astrophysics decadal survey in the
future~\cite{fn}. Since the observational phenomena of pulsars and
compact radioactive source with high energy are often related to
the nuclear processes and the composition of dense matter in
compact stars, it offers us an effective method to probe the
signal of matter composition at super-nuclear density. On one
hand, the equation of state (EOS) and dynamic properties would
affect the structure and evolution behavior of compact stars, on
the other hand, we may estimate the correlative information of
nuclear matter according to the observational macroscopic compact
star properties such as the structure and evolution behavior.
Hence, many investigators spontaneously expect to probe the
super-dense matter through astrophysical observations~\cite{all}.
However, up to now, constraints of inferred masses and radii for
pulsars on equations of state cannot uniquely examine and
distinguish the compositions inside neutron stars. Along with the
abundance of data about spin frequency for fast rotating pulsars,
researchers begin to pay much attention to confirming limit spin
of pulsars, which is another way to probe into the interior of
compact stars.

\section{Method and Result}

For a uniform rigid compact star with mass M and radius R, there
exist some limits on the attainable rotation frequency. The most
obvious one is the Keplerian limit~\cite{Lattimer}
\begin{equation}
\nu_{K}=1.042\times10^{3}\times\frac{(M/M_{\bigodot})^{1/2}}{(R/10km)^{3/2}}
\textmd{Hz}, \label{KC}
\end{equation}
Which is also called the mass-shedding limit. It is roughly
independent of the equation of state. If we get the frequency of
observed pulsar and assume it to be the Keplerian limit, we can
obtain a critical mass-radius relation in the mass-radius plane to
constrain various equations of state for compact stars. This is
the universal treatments as rotation limit by many investigators
which seems to have no help in distinguishing the composition of
super-nuclear matter.

Actually, the emission of gravitational radiation following the
excitation of non-radial oscillation modes may lead to the
instability of rotating stars~\cite{CA}, which can be obtained via
\begin{equation}
\frac{1}{\tau_{G}}+\frac{1}{\tau_{\nu}}=0.
\end{equation}
Here $\tau_{\nu}$ is the damping timescales due to shear, bulk
viscosities and other rubbings, which relates to the viscosities
of the matter inside neutron stars and their differences could
result in diverse behaviors. Based on many works that a star may
spend much more time at the lowest point of r-mode instability
window, we can assume it to be another critical limit on the
attainable rotation frequency. So the genuine limit rotation of
compact stars should actually be decided by both Keplerian and
r-mode constraints. Since we have synthesized both the equations
of state, dynamic properties of matter, this could exert
significantly more stringent constraints to estimate the matter
composition of compact stars.
\begin{figure}[htb]
\begin{center}
\epsfig{file=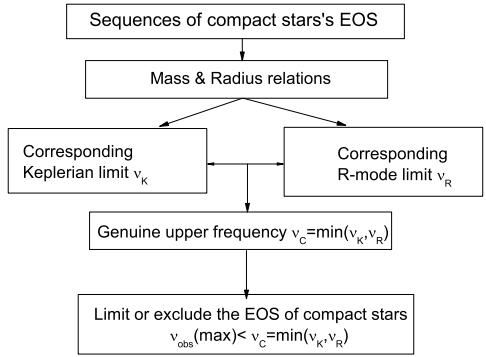,height=2.5in} \caption{Schematic diagram of
thought about our treatment.} \label{fig:1}
\end{center}
\end{figure}

We have used our method shown in Fig 1~\cite{p} to discuss two
rapidly rotating  compact stars. One is SAX J1808.4-3658 with
millisecond rotation period, and the other is XTE J1739-285 with
possible sub-millisecond.
\begin{figure}[htb]
\begin{minipage}[t]{0.5\linewidth}
\centering
\includegraphics[width=3.1in]{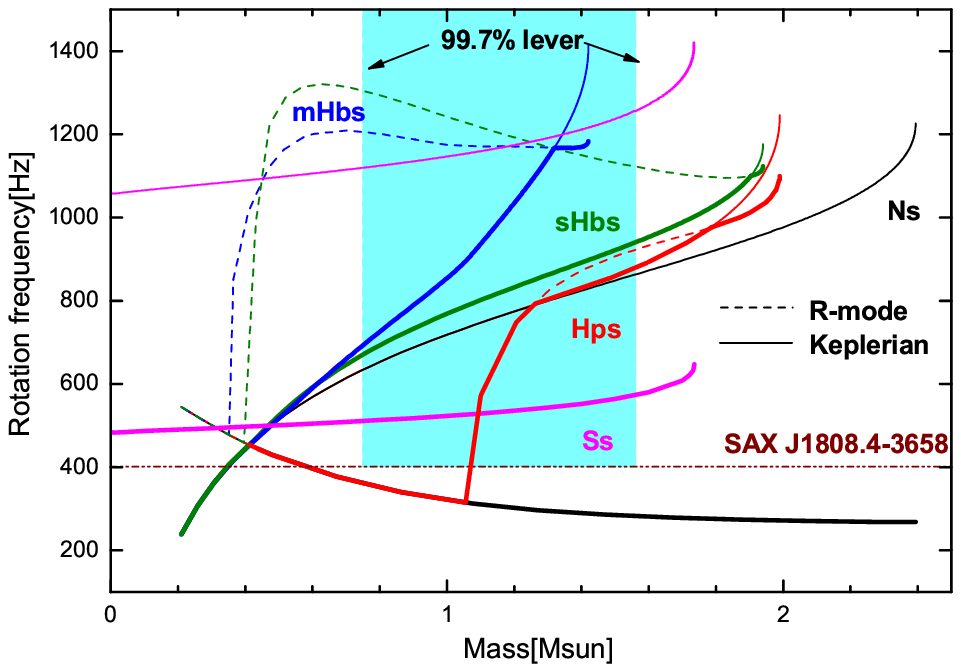}
 \label{fig:side:a}
\end{minipage}%
\begin{minipage}[t]{0.5\linewidth}
\centering
\includegraphics[width=3.1in]{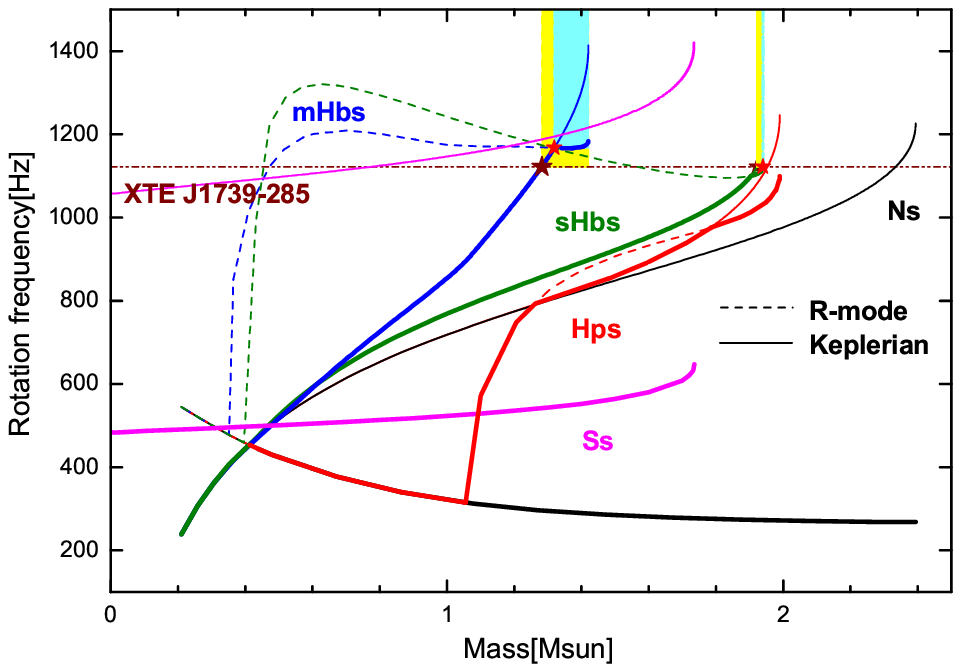}
 \label{fig:side:b}
\end{minipage}
\caption{Limit rotation frequencies for various compact stars. Ns,
Ss, Hps, mHbs and sHbs represent the normal neutron, strange,
hyperon, hybrid star with mixed phase or not respectively. These
corresponding thick solid lines are genuine upper frequencies. In
the left panel,the wine dash-dot-dot horizontal line represents
the 401Hz rotation frequency of the millisecond pulsar in the
transient X-ray burster SAX J1808.4-3658 discovered by Wijnands
and van der Klis ~\cite{wd}; while in the right panel, it
represents the 1122 Hz of burst oscillation in the X-ray transient
XTE J1739-285 that is interpreted as due to the rotation of the
central neutron star by Kaaret et al.~\cite{ka} .}
\end{figure}
In the left panel of Fig 2, the light cyan district refers to our
predicted compatible models that could satisfy the genuine
rotation constraint together with the mass prediction analyzed by
Leahy et al.~\cite{leahy} at 99.7$\%$ confidence lever. We find
that the ordinary neutron star should be excluded. While hyperon
star, hybrid star and strange star are supposed to be the best
candidates for the millisecond pulsar in SAX J1808.4-3658.
Therefore, the star could contain exotic matter either hyperon or
strange quark matter at super-nuclear region, but we could not
tell them from each other, which must depend on more observational
information such as the thermal emission data. However, in the
right panel of Fig 2, we find that only some parts of hybrid star
sequences could spin above 1122Hz. Therefore, we could conclude
that 1122Hz rotation is an obvious evidence for the existence of
quark matter inside the neutron star of XTE J1739-285, or this
possible sub-millisecond pulsar is a hybrid star. Meanwhile,
compared with situation under SAX J1808.4-3658, the compatible
district here is so small that it probably means that the
possibility for the existence of compact stars with
sub-millisecond period is small.

\section{Conclusions}
The composition of matter in the core of neutron stars has
attracted much attention owing to its important significance. In
our estimation, we try to probe the inner components of rapidly
rotating compact stars such as the millisecond pulsar SAX
J1808.4-3658 and the possible sub-millisecond pulsar XTE J1739-285
in our own way by comparing the genuine rotation frequencies under
different theoretical models with the observational data, which
may exert more stringent constraint on matter composition of
compact stars. According to our treatment, the SAX J1808.4-3658 is
a star with exotic matter and XTE J1739-285 a hybrid star.

\section{Acknowledgements}
This work is supported by the National Natural Science Foundation
of China under Grant Nos. 10773004 and 10603002, and the project
A2008-58 supported by the Scientific Research Foundation of
Chongqing University of Posts and Telecommunications.

\def\Discussion{
\setlength{\parskip}{0.3cm}\setlength{\parindent}{0.0cm}
     \bigskip\bigskip      {\Large {\bf Discussion}} \bigskip}
\def\speaker#1{{\bf #1:}\ }
\def\endDiscussion{}

\end{document}

%% file: econfmacros.tex
\def\beq{\begin{equation}}
\def\eeq#1{\label{#1}\end{equation}}
\def\eeqn{\end{equation}}

\def\beqa{\begin{eqnarray}}
\def\eeqa#1{\label{#1}\end{eqnarray}}
\def\eeqan{\end{eqnarray}}

\let\bar=\overbar

\def\Dslash{\not{\hbox{\kern-4pt $D$}}}
\def\dslash{\not{\hbox{\kern-2pt $\del$}}}

\def\msb{{\bar{\ssstyle M \kern -1pt S}}}

\usepackage{fancyhdr,graphicx}